# Learning 3D Mineral Prospectivity from 3D Geological Models Using Convolutional Neural Networks: Application to a Structure-controlled Hydrothermal Gold Deposit

Hao Deng [a], Yang Zheng [a], Jin Chen [a]*, Shuyan Yu [a], Keyan Xiao [b] and Xiancheng Mao [a]

[a] Key Laboratory of Metallogenic Prediction of Nonferrous Metals and Geological Environment Monitoring (MOE), School of Geosciences and Info-Physics, Central South University, Changsha 410083, China

[b] MNR Laboratory of Metallogeny and Mineral Resource Assessment, Institute of Mineral Resources, Chinese Academy of Geological Sciences, Beijing 100037, China

*Corresponding author: chenjingis@csu.edu.cn

Authorship contribution statement

Hao Deng: conceptualization, methodology, software, and writing. Yang Zheng: methodology, software, formal analysis, and writing—original draft. Jin Chen: investigation, resources, and writing—review and editing. Shuyan Yu: formal analysis and visualization. Keyan Xiao: resources. Xiancheng Mao: conceptualization and supervision.

# ABSTRACT

Three-dimensional (3D) geological models are typical data sources in 3D mineral prospectivity modeling. However, identifying prospectivity-informative predictor variables from 3D geological models is a challenging and work-intensive task. Motivated by the ability of convolutional neural networks (CNNs) to learn intrinsic features, in this paper, we present a novel method that leverages CNNs to learn 3D mineral prospectivity from 3D geological models. By exploiting this learning ability, the proposed method simplifies the complex correlations of mineralization and circumvent the need for designing the predictor variables. Specifically, to analyze unstructured 3D geological models using CNNs—whose inputs should be structured—we develop a 2D CNN framework where the geometry of geological boundary is compiled and reorganized into multi-channel images and fed into the CNN. This ensures the effective and efficient training of the CNN while facilitating the representation of mineralization control. The presented method is applied to a typical structure-controlled hydrothermal deposit, the Dayingezhuang gold deposit in eastern China; the presented method is compared with prospectivity modeling methods using designed predictor variables. The results show that the presented method has a performance boost in terms of the 3D prospectivity modeling and decreases the workload and prospecting risk in the prediction of deep-seated orebodies.

1. Introduction

With significant recent advancements in 3D GIS and 3D geological modeling techniques, 3D mineral prospectivity modeling (MPM) has been widely applied in the field of deep prospecting. The 3D MPM, similar to the 2D MPM, comprises three stages of operation (Carranza, 2011; Porwal and Carranza, 2015): 1) the conceptual modeling of the targeted deposits, 2) the derivation of the appropriate predictor variables, and 3) the integration of the predictor variables to output the prospectivity. Among these stages, deriving the mineralization-associated predictor variables is the most critical stage in the MPM process (Carranza, 2011; Porwal and Carranza, 2015).

In contrast to the 2D MPM, one of the most important data sources for 3D MPM is 3D geological models (referred to as 3D models hereinafter, for brevity). The geometry of the geological boundaries potentially indicates the areas with high permeability (Snow, 1969; Lisle, 1994; Liu et al., 2012), the locations of fluid conduits (Kyne et al., 2019; Wilson et al., 2016), and the transition zone of the physical and chemical properties (Ord et al., 2002; Hu et al., 2020; Cao et al., 2020). 3D models are considered to be key data sources for deriving predictor variables in 3D MPM (Wang et al., 2011; Nielsen et al., 2015; Xiao et al., 2015). To extract informative predictor variables from 3D models, intensive research has been conducted on the spatial analysis of 3D models. Spatial analysis methods—such as curvatures analysis (Mejía-Herrera et al., 2015; Li et al., 2015b; Zhang et al., 2019), mathematical morphological analysis (Mao et al., 2016, 2019; Li et al., 2019), lift/depression analysis (Hu et al., 2018; Zhang et al., 2019), geometry transitions analysis (Mao et al., 2019; Liu et al., 2021), roughness analysis (Li et al., 2015), and 3D buffer zones (Li et al., 2016)—were applied predictor variables for 3D MPM. Generally, extracting mineralization-associated predictor variables from 3D models requires considerable domain expertise and a careful trial-and-error process. In addition, the integration of such "hand-crafted" predictor variables can still be limited in terms of representing the underlying correlation to the mineralization and guiding mineral exploration targeting.

Recently, the advent of deep learning (LeCun et al., 2015) has significantly improved many fields of study (Krizhevsky et al., 2012; Hinton et al., 2012; Silver et al., 2016; Esteva et al., 2017; Maxmen, 2018). Deep learning employs deep neural networks to automatically extract informative features from the data, referred to as high-level representation, which helps ascertain intricate patterns and complex associations within the data. The deep-learning approach has been introduced into the field of MPM in recent years (Zuo, 2020; Zuo et al., 2021). To determine their complex relationships with prospective areas, deep neural networks—such as deep autoencoders (DAE) (Xiong et al., 2018; Xiong and Zuo, 2020), convolutional neural networks (CNNs) (Li et al., 2020, 2021a, 2021b; Sun et al., 2020; Xiong and Zuo, 2021; Yang et al., 2021; Zhang et al., 2021), recurrent neural networks (RNNs) (Singh et al., 2018), long short-term memory networks (LSTMs) (Wang and Zuo, 2022), deep regression networks (Xu et al., 2021), variational autoencoder (VAEs) (Luo et al., 2020; Xiong et al., 2021), gated recurrent units (GRUs) (Yin et al., 2021), and generative adversarial networks (GANs) (Zhang and Zuo, 2021; Luo et al., 2021)—have been designed to extract high-level representations from predictor layers in MPM. Most of these deep-learning methods were tailored toward 2D mineral prospecting. While they have led to significant advancements in 2D MPM, to the best of our knowledge, few methods (Li et al., 2021b) have focused on 3D MPM through deep learning.

Most research into 3D MPM focuses on the deposit- and/or camp-scale, which have generally experienced years or decades of exploration and mining. Thus, it is possible to aggregate a large amount of geoscience data with 3D coordinates. Given this large dataset, deep learning is expected to automatically learn a high-level representation associated with mineralization. Li et al. (2021b) proposed the use of 3D CNNs to build a 3D prospectivity model from 3D predictor layers. Despite their methods obtaining promising results, 3D CNN learns from the aforementioned hand-crafted predictor variables derived from conceptual model and spatial analysis, which is not ideal to leverage the strength of deep learning: the capability to automatically extract high-level representations from the primary data close to their raw form. Therefore, we aim to exploit deep networks to learn directly from 3D models and, thus, circumvent the need for the time-consuming process of "hand-crafting" predictor variables.

In this paper, we propose a novel deep-learning method for learning 3D mineral prospectivity from 3D models, with the aim of providing valuable reference material for 3D prospecting. Among the various deep-learning architectures, CNNs have achieved significant success in identifying spatial features in images (Krizhevsky et al., 2012; Sermanet et al., 2014), videos (Karpathy, et al., 2014; Tran et al., 2015), graphical patterns (Lun et al., 2017), and 3D scenes (Wang et al., 2018). Herein, we leverage CNNs to exploit spatial information within 3D models. However, using CNNs to build 3D prospectivity models is a nontrivial task. The challenges are three-fold: firstly, the non-Euclidian and unstructured nature of 3D models prohibits their direct input into CNNs, which require a regular raster input. Secondly, the deep-learning framework should be designed carefully, such that the CNN can learn mineralization-associated features from 3D models. Thirdly, the training of CNNs should be simplified, despite the fact that the training data can be relatively sufficient in 3D MPM. To achieve 3D MPM with CNNs, we reorganize the 3D geological data into 2D multi-channel images for learning the high-level representation from the given 3D models. This facilitates the learning of a detailed geometrical representation of the geological boundaries while allowing us to use mature CNN architectures to build the prospectivity model in an effective and efficient fashion. The prospectivity-informative representation learned using the CNN is then used to output the posterior probability of mineralization. The presented method is applied to a structure-controlled hydrothermal gold deposit, the Dayingezhuang orogenic gold deposit in eastern China, where we have previously extracted "hand-crafted" predictor variables from 3D models and realized 3D MPM (Mao et al., 2019). We show that the prospectivity model—using CNN—outperforms the conventional models using "hand-crafted" predictor variables.

## 2. Methodology

### 2.1. Framework

Given that 3D models are a typically unorganized dataset, an intuitive and natural solution involves voxelizing the 3D geological space and constructing a 3D CNN that directly works in this voxelized 3D space. While 3D CNNs seem to logically be adaptable to 3D models, the parameter sizes of 3D CNNs increase cubically with the 3D convolutional kernel sizes. Therefore, a high-resolution voxel representation that needs larger kernel sizes would incur a large training sample requirement and a prohibitive computational cost to train the CNN. To train a 3D CNN with limited training samples, low-resolution voxels must be used for representing 3D models, which would—in turn—lead to the omission of spatial information in 3D MPM.

Instead, we leverage 2D CNNs to learn the prospectivity informative features by projecting the 3D models onto 2D images, where the 3D information is encoded as multiple channels of the image. This scheme is based on the observation that, despite a 3D model being embedded in 3D space, its boundary surface—which represents the geometry of the geological boundary—is topologically a 2D manifold surface. This motivated the adoption of a 2D CNN to explore the boundary surface of the 3D model. The use of 2D CNNs can ensure the representation of the boundary surface of 3D models in high-resolution while maintaining a smaller training set requirement. Another advantage of using 2D CNNs is that we can re-purpose the available 2D CNN architectures, which have achieved remarkable success in many fields of study (Donahue et al., 2014). Therefore, we realize 3D MPM using 2D CNNs.

We developed a pipeline for this purpose, which is shown in Figure 1. Firstly, the original shape of the 3D models is compiled into several concise but informative shape descriptors (Subsection 2.2). Next, the extracted shape descriptors are projected into multi-channel images (Subsection 2.3) for every target voxel, which encodes the shape of the 3D geological boundary and represents the mineralization control of the target voxel. The projected multi-channel image of target voxels is further fed into the 2D CNN (Subsection 2.4). The 2D CNN transforms the image into a compact high-level representation at the top of the network and outputs the posterior probability of mineralization for the target voxel. In this manner, the CNN associates the 3D models with the ore-bearing probability of the target voxels, resulting in a 3D prospectivity model that does not require the predictor variables to be designed. Further, the workflow facilitates the representation of mineralization control from geological boundaries (Subsection 2.5).

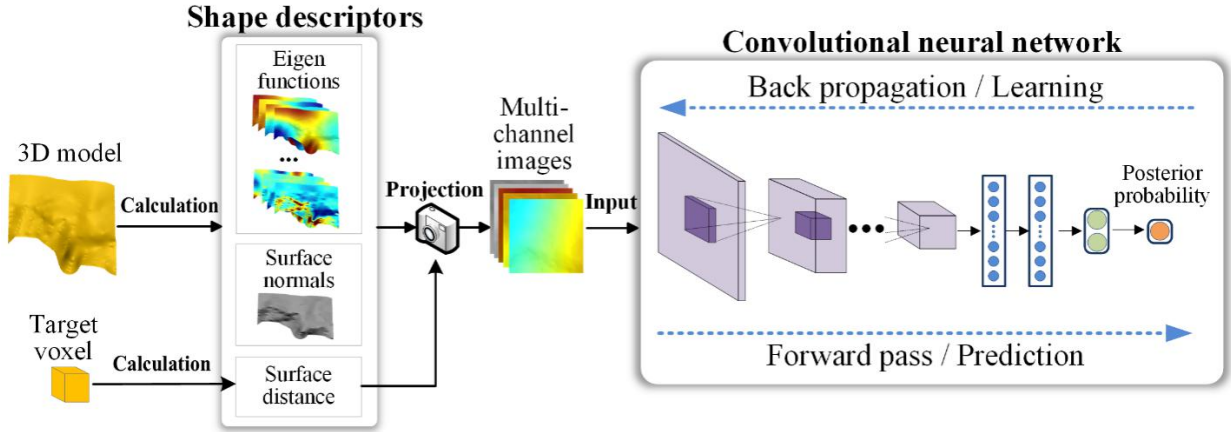

Figure 1: CNN framework for 3D MPM. Given a 3D model, we compile its geometry into several shape descriptors, consisting of Laplace–Beltrami eigenfunctions, shape normals, and the surface distance to each target voxel. The shape descriptors are projected onto multi-channel images for each target voxel via the shape-projection process. The multi-channel images are fed into the convolutional neural network, which learns/predicts the mineral prospectivity for target voxels.

2.2. Shape descriptors

We compile the geometry of the 3D model into several shape descriptors. To fully exploit the available 3D information, we use three types of attributes as the shape descriptors: Laplace–Beltrami eigenfunctions, surface normals, and surface distances. The Laplace–Beltrami eigenfunctions measure the intrinsic structures in the 3D geometry; the surface normal describes the local orientation of the surface; and the surface distance represents the proximity of a given voxel to the geological boundaries.

The Laplace–Beltrami eigenfunctions are derived from the Laplace–Beltrami operator, which is intrinsic to the surface geometry of 3D models (Rustamov, 2007). Representing a real function defined on a compact manifold surface $S$ as $f: M \rightarrow \mathbb{R}$, the Laplace–Beltrami differential operator $\Delta$ over $S$ is defined as the divergence of the gradient:

$$\Delta f = \nabla \cdot (\nabla f), \tag{1}$$

where $\nabla \cdot$ and $\nabla$ denote the divergence and gradient operators over $S$, respectively. Since the Laplace–Beltrami operator is bounded and symmetric positive semidefinite, the eigendecomposition of $\Delta$ can be solved using an eigenproblem:

$$\Delta \phi = \lambda \phi. \tag{2}$$

The solution of Equation (2) results in a series of nonnegative eigenvalues $\lambda_i \geq \cdots \geq \lambda_1 \geq \lambda_0 = 0$ and the associated eigenfunctions $\phi_i$, which are referred to as the Laplace–Beltrami eigenfunctions.

The Laplace–Beltrami eigenfunctions form the Fourier basis for the Fourier analysis on manifold surfaces (Hammond et al., 2011). Figure 2 illustrates the first 16 Laplace–Beltrami eigenfunctions for a simple cube-like surface. It can be seen that these eigenfunctions correspond to low-frequency wave functions that conform to the surface geometry. Laplace–Beltrami eigenfunctions offer two important characteristics in our framework:

1) The eigenfunctions are intrinsic to the surface shape. The values of the eigenfunctions defined on the surface are independent of the Cartesian coordinates of the surface, but are dependent on the Laplace–Beltrami operator in terms of the intrinsic geometry of the surface. The eigenfunctions are invariant under isometric deformation due to the isometric invariance property of the Laplace–Beltrami operator. Thus, the eigenfunctions characterize the intrinsic geometry and are referred to as the "shape-DNA" (Reuter et al., 2006) of the surface.

2) The eigenfunctions form a complete basis for the space of smooth functions on the surface. The eigenfunctions are orthogonal to each other over the manifold surface $S$. That is, the inner product of the eigenfunctions $\phi_i$ and $\phi_j$ satisfies

$$\langle \phi_i, \phi_j \rangle = \int_S \phi_i \phi_j = \begin{cases} 0, & i \neq j, \\ 1, & i = j. \end{cases} \tag{3}$$

Thus, an arbitrary square-integrable function on the surface can be expanded using the eigenfunctions:

$$f = a_1 \phi_1 + a_2 \phi_2 + \cdots, \quad a_i = \langle f, \phi_i \rangle \tag{4}$$

Equation (4) shows that the eigenfunctions allow us to approximate the underlying functions on the surface of the geological boundary to represent mineralization controls in our framework. We will justify this in Subsection 2.5.

The discretized Laplace–Beltrami eigenfunctions for 3D models are based on the discretization of the Laplace–Beltrami operator. We adopt the approach proposed by Rustamov (2007) to compute the discretized versions of the Laplace–Beltrami eigenfunctions at every vertex of the 3D models. We refer interested readers to Zhang et al. (2010) for more details regarding Laplace–Beltrami eigenfunctions.

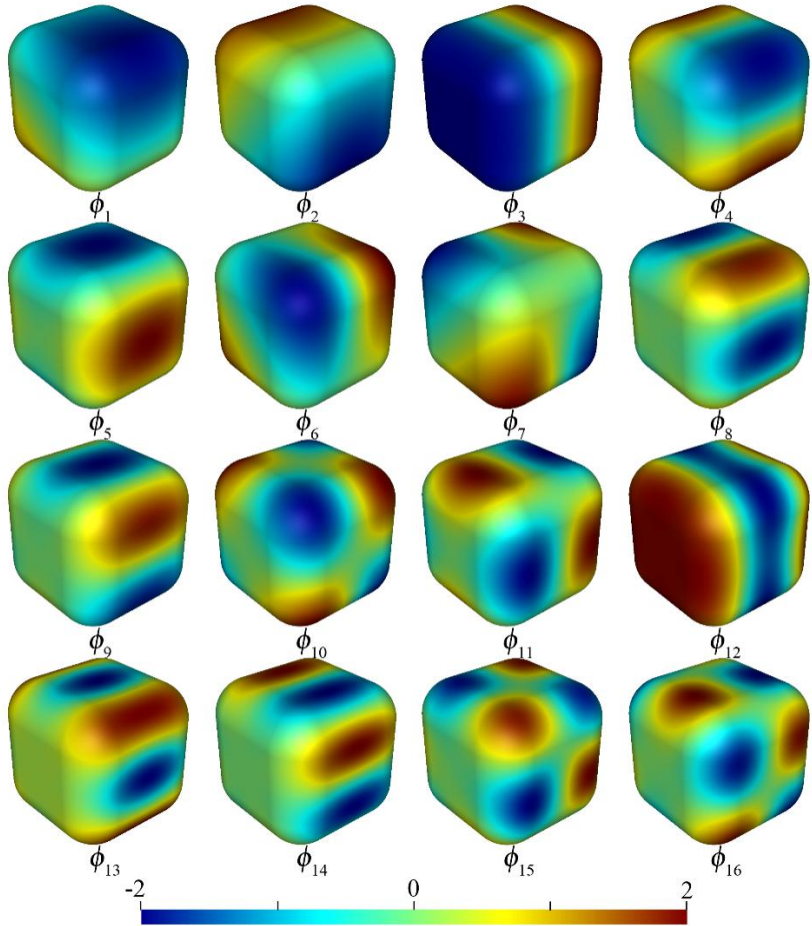

Figure 2: Laplace–Beltrami eigenfunctions for a cube-like surface. Note that the eigenfunctions conform to the geometric features of the surface.

2.3. Shape projection

We reorganize the shape descriptors into multi-channel images through shape projection, where each channel corresponds to a single descriptor. Given a target voxel, the shape projection generates a projected image that encodes the shape of the geological boundary associated with the given voxel. We use the perspective projection to project shape descriptors onto images. A characteristic of prospective projection is that a farther face has a smaller projected area than a closer one. Therefore, in the shape projection, the projected area represents the diminishing mineralization control with increases in the distance to the geological boundary.

To ensure the diversity of training data and high-resolution projection, in the projected image, we only capture a local area of the geological boundary "viewed" from the target voxel. Thus, for the target voxel, it is essential to capture its major control area on the geological boundary. The captured area in the projected image depends on the "viewing direction" from the target voxel to the geological boundary. This means searching for the major control area can be seen as a viewing-direction search problem.

To formulate the viewing-direction search problem, we first estimate the mineralization control from the area captured by the viewing direction. For a target voxel, represented by its center $\mathbf{v}$, and a 3D model of the geological boundary, represented by its surface $S$, if we project $S$ through the viewing direction $\mathbf{d}$, a captured area $\text{CapArea}(\mathbf{d}, \mathbf{v}) \subset S$ on the geological boundary can be generated. The mineralization control from $\text{CapArea}(\mathbf{d}, \mathbf{v})$ to $\mathbf{v}$ can be estimated by the surface integral over $\text{CapArea}(\mathbf{d}, \mathbf{v})$:

$$E_{\mathbf{v}}(\mathbf{d}) = \int_{\text{CapArea}(\mathbf{d}, \mathbf{v})} C(\mathbf{x}, \mathbf{v}) d\mathbf{x}, \tag{5}$$

where $C(\mathbf{x}, \mathbf{v})$ denotes the mineralization control from a point $\mathbf{x}$, on $S$, to $\mathbf{v}$. The definition of $C(\mathbf{x}_i, \mathbf{v})$ in Equation (5) is essential to estimate $E_{\mathbf{v}}(\mathbf{d})$. Considering that mineralization control is closely related to the proximity to the geological boundary, and the fact that some previous studies (Carranza and Hale, 2002; Nielsen et al., 2015; Li et al., 2015, 2016) have used the distance to represent geological control, in our model, we use the inverse distance $C(\mathbf{x}, \mathbf{v}) = 1/\|\mathbf{x} - \mathbf{v}\|$ to measure the mineralization control from $\mathbf{x} \in S$ to $\mathbf{v}$. We will justify, in subsection 2.5, that this metric approximates the heat/mass transport controlled by geological boundaries.

In terms of the above definition, the major control area for $\mathbf{v}$ can be defined as a captured area $\text{CapArea}(\mathbf{d}^*, \mathbf{v})$ on the geological boundary captured by an optimized viewing direction $\mathbf{d}^*$, such that the mineralization control $E_{\mathbf{v}}(\mathbf{d})$ from $\text{CapArea}(\mathbf{d}^*, \mathbf{v})$ to $\mathbf{v}$ is maximized. Thus, the viewing direction is searched for by solving an optimization problem, which results in an optimized viewing direction $\mathbf{d}^*$ to capture the major control area. Thus,

$$\mathbf{d}^* = \arg\max_{\mathbf{d}} E_{\mathbf{v}}(\mathbf{d}). \tag{6}$$

The optimization problem in Equation (6) is highly nonlinear. Using the gradient-based optimization algorithm to solve Equation (6) is difficult since estimation of the gradient is intractable. Inspired by the mean-shift algorithm (Comaniciu and Meer, 2002), we design an algorithm—summarized in Algorithm 1—to solve the optimization problem. Initially, we set $\mathbf{d}$ towards the closest point $\mathbf{c}_0 \in S$ to $\mathbf{v}$. Next, we generate a multi-channel image with $\mathbf{d}$ and $\mathbf{v}$. The centroid $u$ of the image to adjust $\mathbf{d}$ is calculated to be weighted by $C(\mathbf{x}, \mathbf{v})$:

$$u = \frac{\sum_{i \in \mathcal{I}} u_i C(\mathbf{x}_i, \mathbf{v})}{\sum_{i \in \mathcal{I}} C(\mathbf{x}_i, \mathbf{v})}, \qquad (7)$$

where $\mathcal{I}$ is the set of pixels corresponding to the projection of $S$, $u_i$ is the image coordinate of the pixel in $\mathcal{I}$, and $\mathbf{x}_i$ is the corresponding point on $S$ projected to $u_i$. Subsequently, we unproject $u$ to determine the corresponding point $\mathbf{c}$ on $S$. Finally, we update the viewing projection $\mathbf{d}$ towards $\mathbf{c}$ to generate a new multi-channel image. The above mean-shift procedure is repeated until the convergence of $\mathbf{d}$. Figure 3 illustrates an example of this optimization.

---
**Algorithm 1** Mean-shift algorithm to solve Equation (6)
---
**Input:** Surface $S$ of the geological boundary, target voxel $\mathbf{v}$
**Output:** Viewing direction $\mathbf{d}$
Set $\mathbf{d}$ towards the closest point $\mathbf{c}_0 \in S$ to $\mathbf{v}$;
**while** not converged **do**
    Multi-channel image $I$ = ShapeProjection($\mathbf{d}$, $\mathbf{v}$);
    Calculate weighted centroid $u$ of $I$ according to Equation (7);
    Unproject $u$ to find the corresponding point $\mathbf{c}$ on $S$;
    Update $\mathbf{d}$ towards $\mathbf{c}$;
**end**
---

2.4. Convolutional neural networks

To extract the features associated with mineralization, we leverage the well-established CNN architecture of AlexNet (Krizhevsky et al., 2012) as the backbone for learning high-level representations from the projected images. As shown in figure 4, our AlexNet-based CNN comprises five convolutional layers (Conv1 to Conv5) and three subsequent fully connected layers (FC6–FC8), from bottom to top. In contrast to AlexNet, in our CNN, the convolutional kernel size at Conv1 is modified from $11 \times 11 \times 3$ to $11 \times 11 \times n$ to fit the $n$-channel images. Moreover, the original 1000-class softmax classifier at the head of the network is replaced by a two-class logistic classifier for outputting the posterior probability of mineralization.

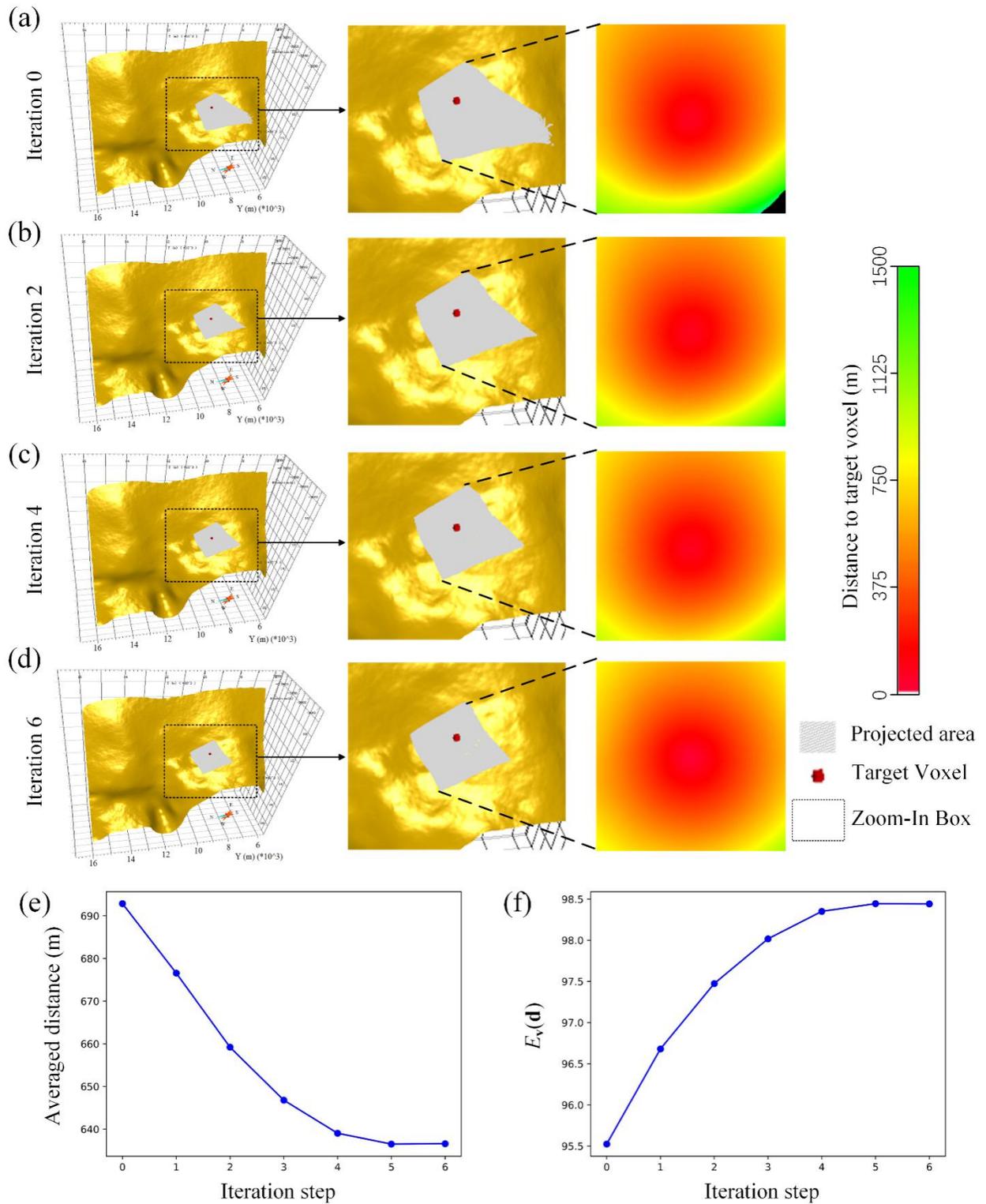

Figure 3: Example of view direction optimization. Given a target voxel and a 3D model of geological boundary, initially, the viewing direction is set towards the closest point to the target voxel at the geological boundary (a),

which causes parts of the distant geological boundary to be projected onto the image (right column of (a)). With the optimization of the viewing direction ((b)–(d)), the projected areas converge closer to the target voxel (right columns of (b)–(d)), which can also be reflected in the variations of the averaged distance to the target voxel (e) and the maximization process of objective functions $E_\mathbf{v}(\mathbf{d})$ (f).

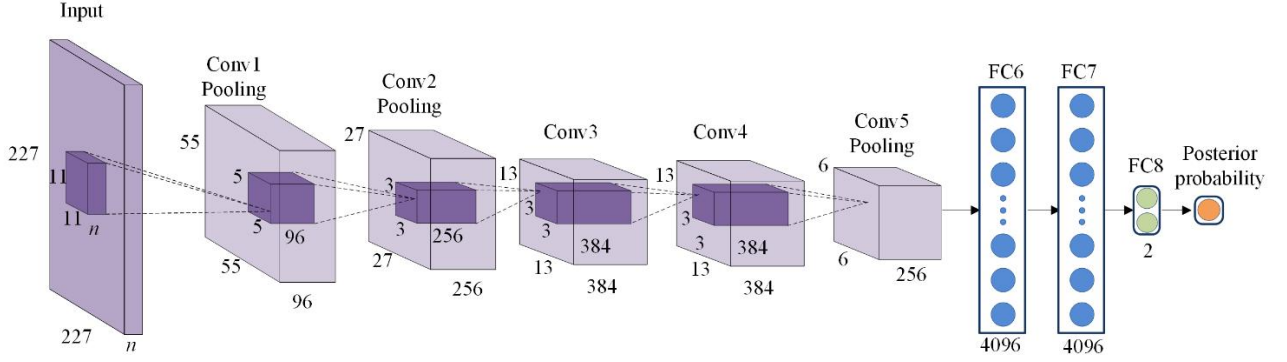

Figure 4: AlexNet-based CNN architecture.

We train the above CNN for 3D MPM by using the information of $n$ known voxels. Each voxel is associated with a multi-channel image $X_i$ and a label $y_i$, with 1 representing "ore-bearing" and 0 representing "non-ore-bearing". We train the parameters $\theta$ of the CNN model that outputs the posterior probability $f_\theta(X)$ of mineralization by minimizing the loss function $L(\theta)$. The objective function includes a cross entropy loss to force the CNN to fit the training data, and a regularization term to prevent the CNN from overfitting:

$$L(\theta) = -\left(\sum_{i \in \mathcal{P}} \log f_\theta(X_i) + \sum_{i \in \mathcal{N}} \log\left(1 - f_\theta(X_i)\right)\right) + \lambda \|\theta\|^2, \tag{8}$$

where $\mathcal{P}$ and $\mathcal{N}$ are the sets of ore-bearing and non-ore-bearing samples, respectively, and $\lambda$ is the weighting for the regularization term.

The class imbalance of training samples is a common issue in MPM (Xiong and Zuo, 2018). Generally, non-ore-bearing samples are the majority class, making the ore-bearing samples an under-represented minority class. This can lead to the under-classification of the ore-bearing class. To mitigate this issue, we can adopt the focal loss (Lin et al., 2017) for training the CNN model, which modifies the focal loss by introducing a modulating factor. Based on the formulation in Equation (8), focal loss is defined as:

$$L(\theta) = -\left(\sum_{i \in \mathcal{P}} \left(1 - f_\theta(X_i)\right)^\gamma \log f_\theta(X_i) + \alpha \sum_{i \in \mathcal{N}} f_\theta(X_i)^\gamma \log\left(1 - f_\theta(X_i)\right)\right) + \lambda \|\theta\|^2, \tag{9}$$

where $(1 - f_\theta(X_i))^\gamma$ and $f_\theta(X_i)^\gamma$ are the modulating factors for the two classes, $\gamma \geq 0$ is the focusing parameter, and $\alpha$ is the weighting factor between the two classes. The modulating factor greatly suppresses the loss contributions from the easily classified samples, which are expected from the majority class, and enhances the relative effect of hard, misclassified samples, which are expected from the minority class. In practice, we used $\gamma = 2$, following the study by Lin et al. (2017), and set $\gamma$ to the ratio of ore-bearing samples and non-ore-bearing samples to balance the contribution from the two classes.

2.5. Justification

The above workflow for 3D MPM can be interpreted from the perspective of ore-forming dynamics. The geological boundaries, such as faults and intrusive contacts in structure-controlled mineral deposits, play a significant role in releasing and driving hydrothermal fluids. The hydrothermal fluid flow is essential for heat and mass transport in permeable rocks, which dominates the formation of favorable areas for metal trapping and deposition in the proximity of geological boundaries. The key governing equation for the heat and mass transport of hydrothermal fluids in a steady-state can be formulated using the convection–diffusion–reaction equation (Zhao et al., 2008):

$$\nabla \cdot (D\nabla C) - \nabla \cdot (Cq) + KC + F = 0, \tag{10}$$

where $C$ is the concentration (for mass transfer) or temperature (for heat transfer) at position $x \in \mathbb{R}^3$ and time $t \in \mathbb{R}$, $D$ is the diffusivity, $q$ is the Darcy velocity, $K$ is the first-order reaction velocity, and $F$ is the source function.

Applying the Green's function method, the solution of Equation (10) has the following form if we only consider the area $V$ enclosed by the geological boundary as the local heat/mass source:

$$C(x) = \iiint_V G(x, y) F(y) dy, \tag{11}$$

where $G(x, y)$ is the Green's function for the partial differential equation in Equation (10). $G(x, y)$ represents the "response" at $y$ to a heat/mass at $x$, and is rather complex in 3D space. Only considering the long-term diffusion by the discarding the convection and reaction phenomena in Equation (10), we have $G(x, y) = \frac{1}{\|x-y\|}$, which is consistent with the inverse distance metric of mineralization control in the shape projection.

Using the divergence theorem, Equation (11) is transformed into

$$C(x) = \oiint_S H(x, y) \cdot \boldsymbol{n} dy, \tag{12}$$

where $H(x, y) = \frac{1}{4\pi} \iiint_V G(x, z) F(z) \frac{y-z}{\|y-z\|^3} dz$, and $\boldsymbol{n}$ is the surface normal of the geological boundary.

Equation (12) demonstrates that the mass/heat distribution controlled by the geological boundary can be expressed by the surface integral over the surface of the geological boundary. Note that our 2D CNN learns from the images of the projected geological boundary. It can be considered to represent the mineralization control by learning to "integrate" over the major control areas of the geological boundary. Recall that the multi-channel images include channels of Laplace–Beltrami eigenfunctions, the linear combination of which is capable of approximating an arbitrary square-integrable function $f$ on the surface. We prove that, in the CNN, the convolution of the eigenfunctions with kernels in the CNN is equivalent to the linear combination of the eigenfunctions in the surface domain (see Appendix I in the supplementary material). This allows the first convolution layer of the CNN to generate a family of functions in the surface domain, to represent mineralization controls in the following layers. Thus, the CNN is expected to derive high-level representations related to mineralization control from the geological boundary, which facilitates building the association between the geological boundary and the mineralization.

Given that the operations in the first convolution layer generate a family of functions on the surface of the geological boundary, our workflow can be also regarded to be consistent with conventional 3D MPM methods, which use spatial analysis to extract shape features on geological boundaries. In contrast to these methods, the CNN model extracts a family of features of geological boundaries as a learning problem, which automatically optimizes the features for 3D MPM tasks in the learning process.

## 3. Study area and data

### 3.1. Geological setting of the study area

Our study area, the large-scale Dayingezhuang Au deposit (> 130 t), is located in the northwestern Jiaodong Peninsula (Figure 5), the largest gold province in China. The deposit is principally hosted in the Zhaoping fault, which exhibits the metamorphosed Archean Jiaodong Group as the hanging wall and Jurassic Linglong granite as the footwall of the Zhaoping fault (Figure 6). The Zhaoping detachment fault, developed along the interface between the Jiaodong Group rocks and the Linglong granite, generally strikes an SW–NE direction, with SE dips of 35°–60° (Mao et al., 2019). It exhibits gentle shape undulation along both the strike and dip directions.

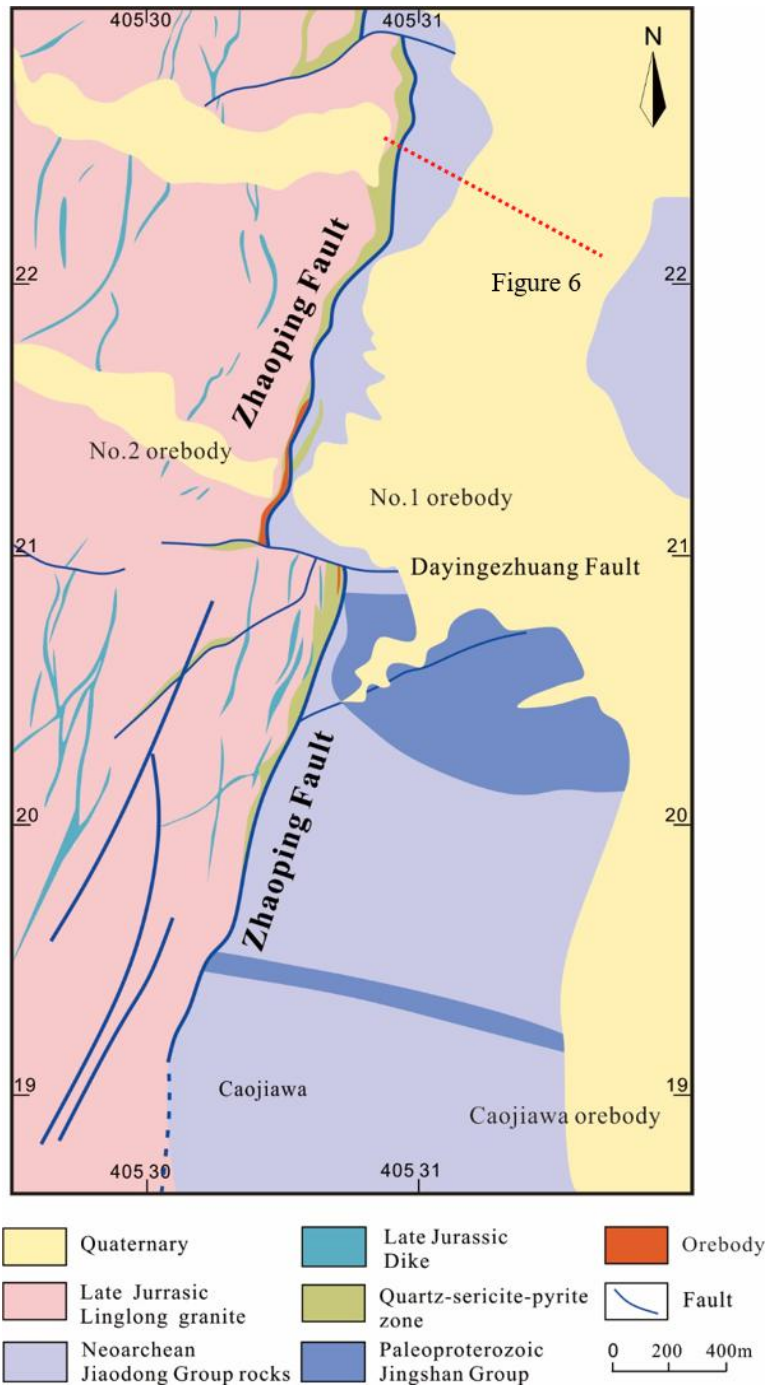

Figure 5: Geological map of the Dayingezhuang gold deposit (adapted from Mao et al., 2019).

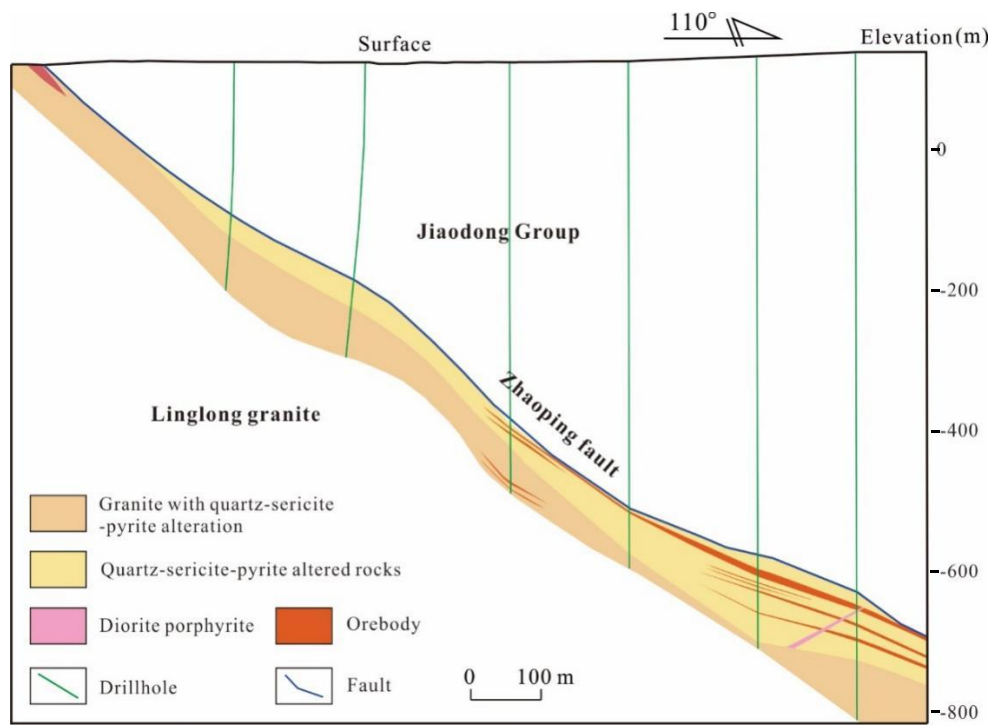

Figure 6: Mineralization-alteration zoning pattern (at section line #86) of the Dayingezhuang gold deposit.

The localization of gold orebodies in Dayingezhuang is closely related to the geometry of the Zhaoping fault. Most Au mineralization is hosted in the mylonization and brecciation zones in the footwall of the Zhaoping fault. There are major three orebodies—i.e., No. 1, No. 2, and Caojiawa orebodies—in the Dayingezhuang deposit. The No. 2 orebody, with Au grades typically < 5 g/t, contains 65% of the Au resources in the deposit. It extends along the strike direction more than 3000 m and 1500 m NE, plunging deep and exhibiting disseminated mineralization and intense quartz–sericite–pyrite alterations. The No. 1 orebody is located in the southern segment of the Dayingezhuang deposit and extends about 1500 m along the strike direction, which consists of disseminated ores and vein ores. The Au grade of the No. 1 orebody ranges from 2–4 g/t. The Caojiawa orebody is located to the south of the No. 1 orebody and extends about 900 m, with Au grades < 2 g/t, which is also characterized by the disseminated mineralization.

Previous studies have demonstrated that the Dayingezhuang gold deposits and other gold deposits (e.g., Jiaojia and Sanshandao) in the Jiaodong Peninsula could be classified as atypical orogenic deposits, i.e., Jiaodong-type gold deposits (Goldfarb and Santosh, 2014; Yang et al., 2014, 2016; Song et al., 2015; Deng et al., 2020; Liu et al., 2021a). These gold deposits have similar structural controls in terms of mineralization and ore-forming conditions (moderate temperatures and salinities, along with $CO_2$-rich fluids). The gold mineralization is closely associated with the regional detachment faults (e.g., dips and geometry), but is not associated with the host rocks or any exposed intrusion (Goldfarb and Santosh, 2014; Li et al., 2015a; Mao et al., 2019). The significant structural controls on mineralization essentially result from the influences of structural deformation, driving fluid flows, trapping ore-forming fluids, and/or metal deposition under fluctuating pressure conditions in fault zones on rock permeability (Yang et al., 2016, 2018; Mao et al., 2019; Wang et al., 2019; Liu et al., 2021b). Owing to the dominated controls from the detachment faults, the location of gold mineralization in Dayingezhuang can be well-inferred from the geometry of the Zhaoping fault (Mao et al., 2019).

### 3.2. Data collection and 3D modeling

To conduct 3D MPM in the Dayingezhuang gold deposit, a current and legacy mine dataset was collected, containing six surface maps, 24 subsurface horizontal sections, 76 cross-sections, 187 drillholes, 56 tunnels, 16,747 gold assays, and 23 magnetotelluric (MT) sounding profiles (until June 2020). First, the 3D models of the Dayingezhuang deposit were constructed using the exploration data. Here, the shape of the Zhaoping fault was delineated following the cross-section profiles. For the areas where the exploration data was scarce or missing, the location of the Zhaoping fault was carefully inferred and delineated according to the MT profiles (Mao et al., 2019). Finally, the implicit modeling method (Macedo et al., 2011) was used to interpolate the geometry between the delineated contours. Finally, the 3D model of the Zhaoping fault was generated from the implicit surface (Figure 7(a)).

To represent the 3D distribution of mineralization, the 3D models of orebodies were constructed from the exploration data (Figure 7(b)). The voxel models representing Au distributions were built in terms of the 3D models of orebodies. Here, the 3D area of Dayingezhuang was divided into regular voxels, each with a size of 25 m × 25 m × 25 m. The kriging method was utilized to interpolate the gold grades for these voxels in terms of Au assay samples. According to the cutoff grade of 1.0g/t gold, 4,997 voxels (Figure 7(c)) were labeled as ore-bearing, while the remaining 21,943 voxels were labeled as non-ore-bearing.

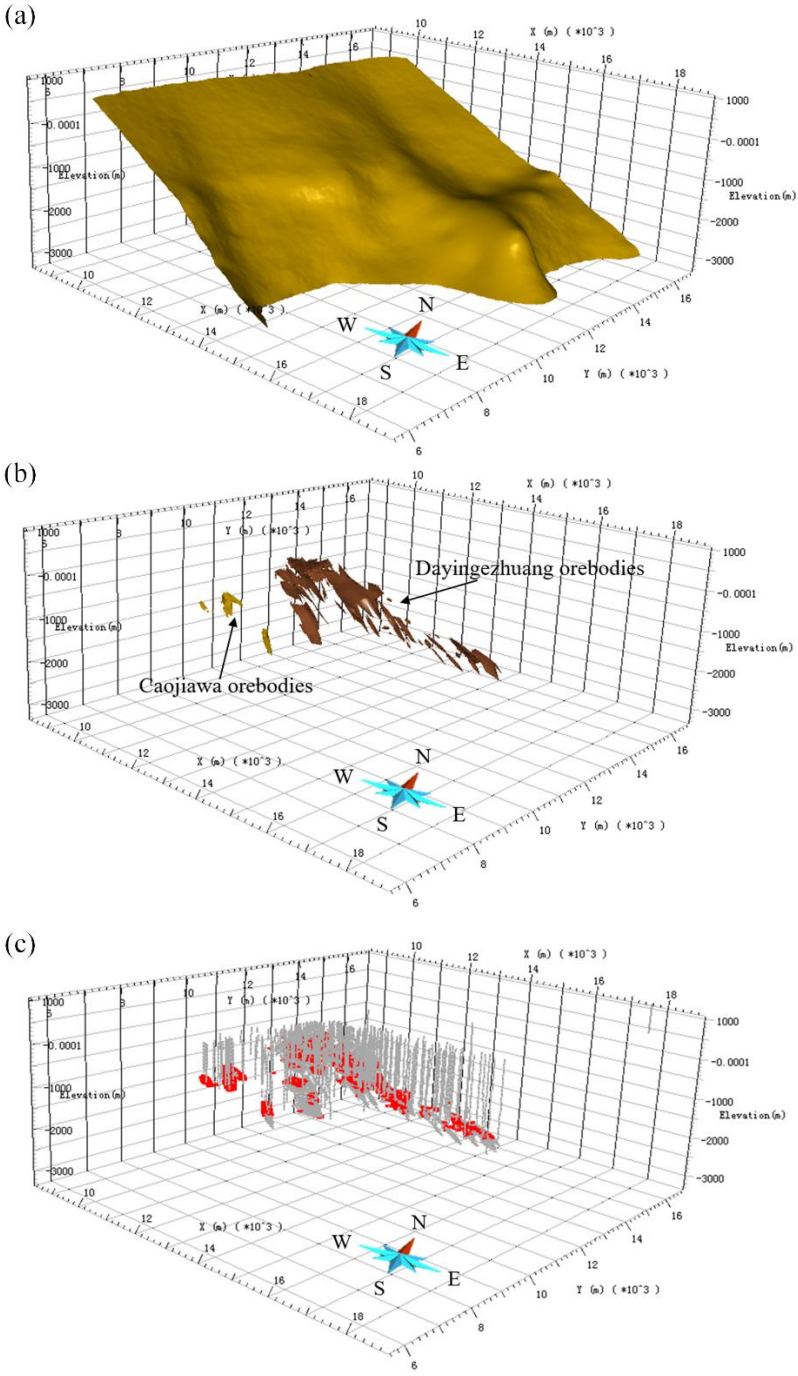

Figure 7: 3D models of the (a) Zhaoping fault, (b) orebodies, and (c) known voxels (Au grade >1g/t).

## 4. Results

### 4.1. 3D MPM with convolutional networks

Firstly, we calculate the shape descriptors of the 3D model of the Zhaoping fault. The first 16 Laplace–Beltrami eigenfunctions with non-zero eigenvalues were used to encode the intrinsic shape of the Zhaoping fault, which—in addition to the surface normals—form the shape descriptors for the 3D model (Figure 8).

To generate the training and testing dataset, for each target voxel, a 20-channel image was projected at a fixed resolution of $227 \times 227 \times 20$, containing 19 channels of shape descriptors, as shown in Figure 9 and one channel of the surface distance to the voxel. For the training of the CNN, a ratio of 8:2 was used to partition known voxels into the training and validation sets.

We implemented our model in TensorFlow (Abadi et al., 2016). The CNN was trained with respect to the training set in a fully supervised fashion. During implementation, the supervised training took 50 epochs to fully converge, requiring about 15 hours on an Nvidia RTX 2080Ti 11GB graphics card. During the training process, the validation accuracy (training loss) increases (decreases) progressively for both the training and validation set (Figure 10). To evaluate how the training impacts the prospectivity of the unknown areas in Dayinggezhuang, the fuzziness (Zadeh, 1965) of the predicted posterior probability was estimated. Similar to Figure 10, the fuzziness of unknown areas gradually decreases during the training (Figure 11). Overall, the aforementioned results demonstrate that the CNNs can be trained to identify the association with mineralization.

As we used a 2D CNN architecture, the CNN could be pre-trained using the ImageNet dataset (Russakovsky et al., 2015), which is a well-established paradigm in deep learning. We compared the validation accuracy with and without ImageNet pre-training. The results (Figure 12) show that the ImageNet pre-training only led to a slight performance improvement. This may be caused by the significant differences between the ImageNet images, which are detailed natural images, and our multi-channel images that represent abstracted spatial structures. However, it can be seen that the ImageNet pre-training speeds up the convergence of the training process (Figure 12). This is probably caused by the fact that the CNN trained without ImageNet pre-training has to learn more low-level features, such as low-frequency variations and textures in low-level CNN layers. Further, the results are also consistent with recent findings in the machine-learning community (He et al., 2019; Zoph et al., 2020). Thus, ImageNet pre-training facilitates rapid prospectivity modeling and makes it easier to gain encouraging results, which is an added benefit of using 2D CNNs.

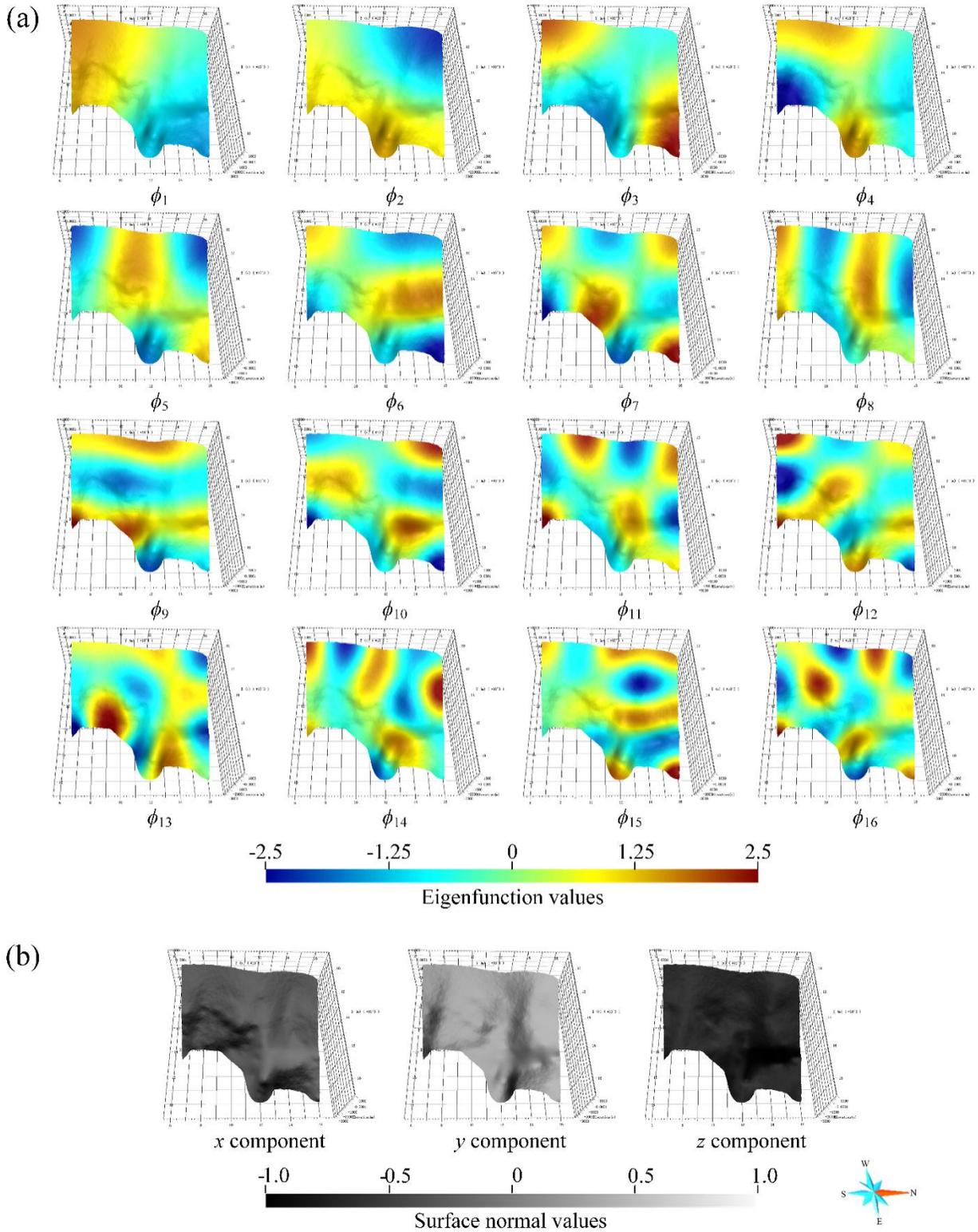

Figure 8: Shape descriptors of Laplace–Beltrami eigenfunctions (a) and surface normals (b) of the Zhaoping fault in the Dayingezhuang segment.

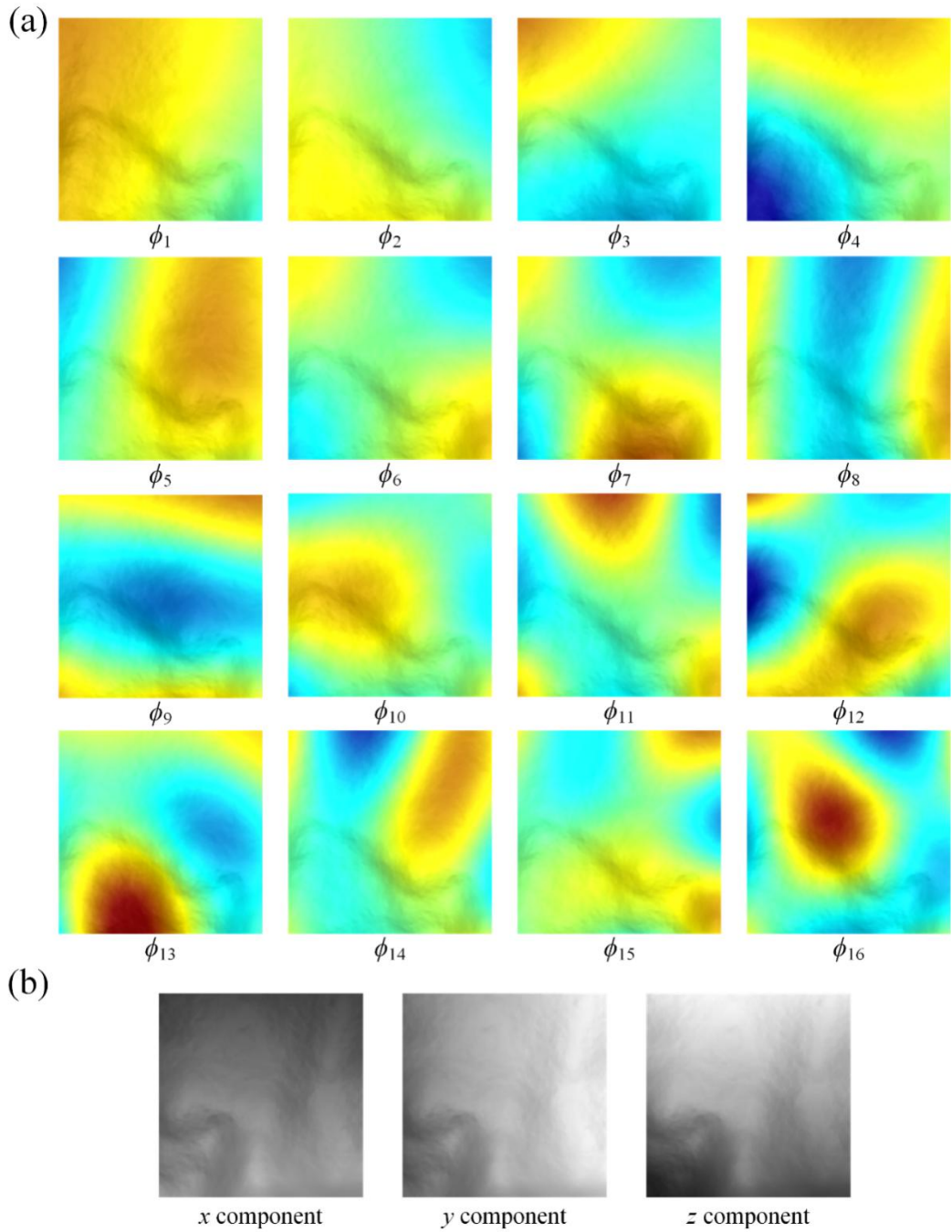

Figure 9: Projection results of the shape descriptors for a target voxel: Laplace–Beltrami eigenfunctions (a) and surface normals (b).

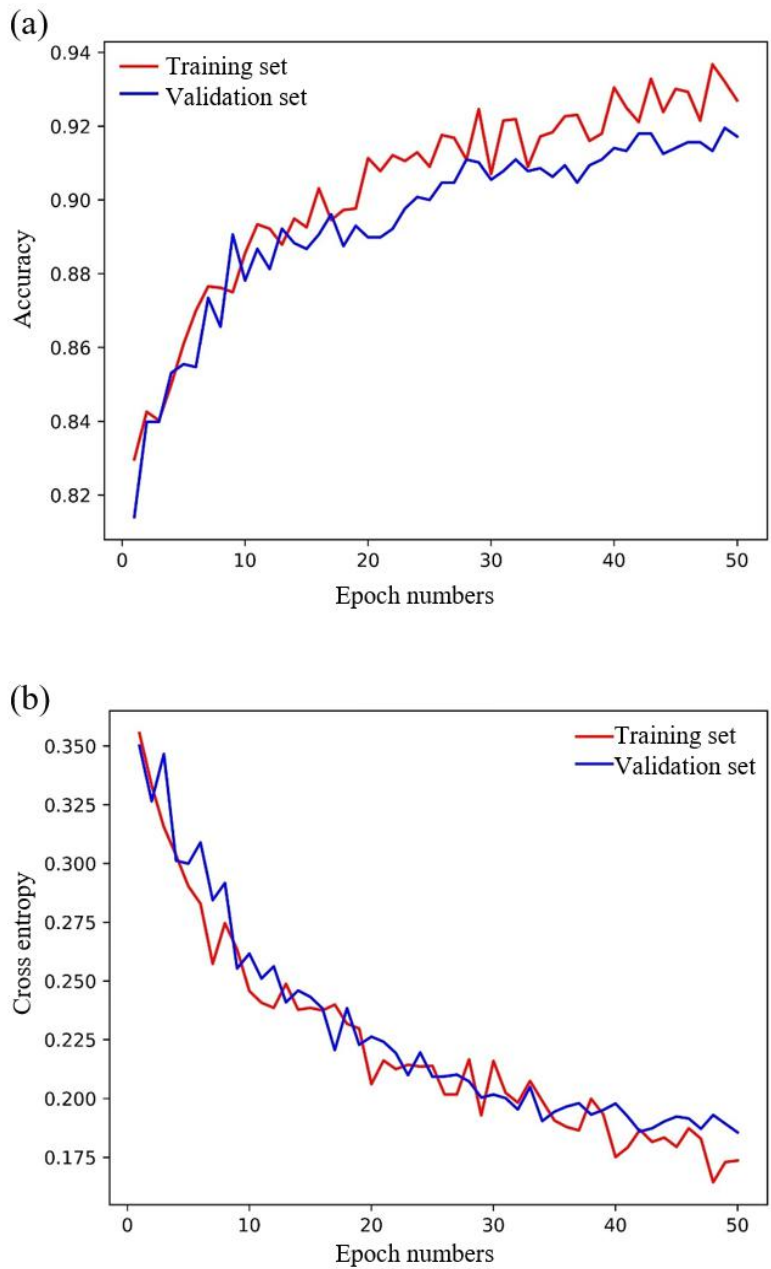

Figure 10: Variation of accuracy (a) and cross entropy loss (b) during the training process.

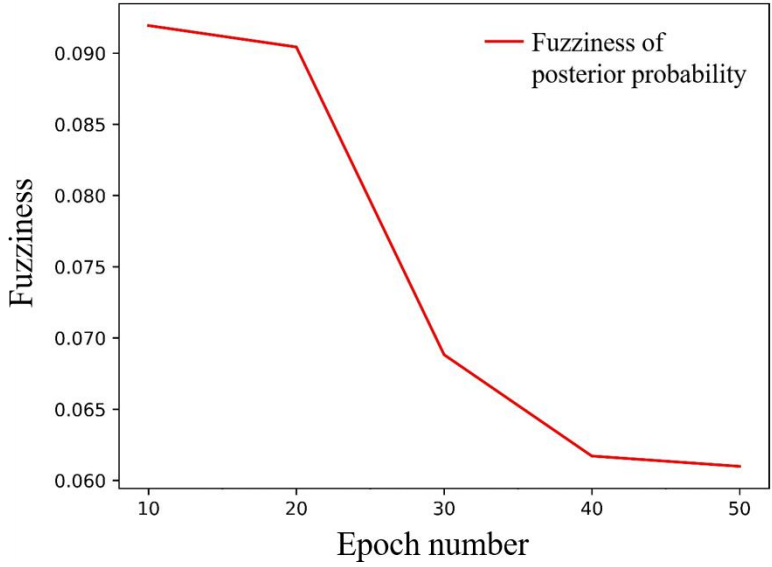

Figure 11: Variation of fuzziness for unknown areas. In a similar manner to the decreased cross entropy measured in known areas, the fuzziness of unknown areas gradually decreases during the training process.

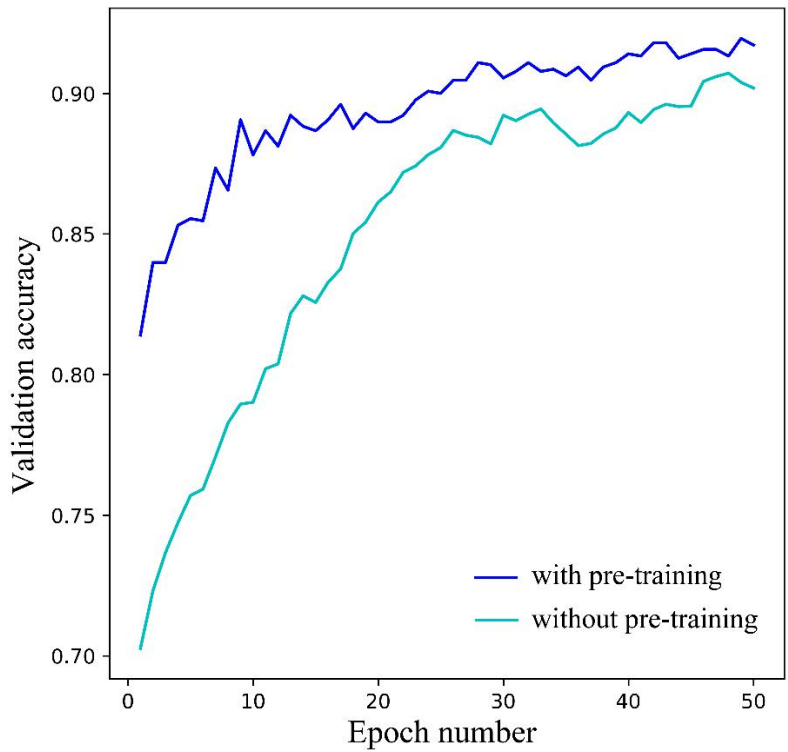

Figure 12. Variation of validation accuracy with and without ImageNet pre-training. The ImageNet pre-training only provides a slight improvement of validation accuracy but leads to an earlier convergence to promising results.

### 4.2. 3D MPM based on hand-crafted predictor variables

To validate the advantages of the proposed method in terms of 3D MPM, we compared our method with the conventional method that uses hand-crafted predictor variables. Here, the hand-crafted predictor variables are the same as those in a previous study by Mao et al. (2019). Table 1 summarizes these predictor variables. The competing prospectivity models were constructed by integrating the predictor variables using machine-learning models. We tested four machine-learning models, including logistic regression (LR) (Agterberg et al., 1993; Li et al., 2015b), support vector machines (SVMs) (Zuo et al., 2011; Ghezelbash et al., 2021), random forests (RFs) algorithms (Carranza and Laborte, 2015; Xiang et al., 2020), and multi-layer perceptrons (MLPs) (Abedi and Norouzi, 2012; Ghezelbash et al., 2020). To train the model, the voxels were also partitioned into the training and validation sets in the same manner as the setting for the CNN model. Further, the hyperparameters of these four models were fine-tuned via 10-fold cross-validation.

Table 1: Targeting criteria, predictor variables, and the corresponding spatial analysis approach for 3D MPM in the Dayingezhuang deposit.

| Targeting criteria | Predictor variable | Spatial analysis approach |
| --- | --- | --- |
| Proximity to the Zhaoping fault | Euclidian distance to the Zhaoping fault | 3D Euclidean distance transform |
| Shape variation of the Zhaoping fault | 1st-order/2nd-order undulation of the Zhaoping fault | 3D morphological undulation analysis |
| Dipping of the Zhaoping fault | Slope of the Zhaoping fault | Slope analysis |
| Dipping transition of the Zhaoping fault | Dipping variation of the Zhaoping fault | 3D geometry transition extraction |
| Hydrothermal alteration | Thickness of the alteration zone | Field analysis of the alteration zone |

### 4.3. Performance evaluation

To evaluate the performance of the prospectivity models, the receiver operating characteristic (ROC) curves (Barreno et al., 2007) and success-rate curves (Agterberg and Bonham-Carter, 2005) were utilized.

The ROC curves were generated by plotting the true positive rate (TPR) against the false positive rate (FPR) at different thresholds for separating the high-prospective areas. The area under the curve (AUC) was used to evaluate the ROC curves. A higher AUC indicates a better performance of the prospectivity model. In our experiment, only the voxels in the validation set were employed to plot the ROC curves. Figure 13(a) illustrates the ROC curves for the prospectivity models. The ROC curve of the CNN model has an AUC value of 0.96, the highest among the five evaluated prospectivity models. This demonstrates that the CNN model outperforms the other four models, which use hand-crafted variables, in the field of 3D MPM.

On the other hand, the success-rate curves were generated by calculating the percentage of ore-bearing voxels against all predicted voxels under the different thresholds. A higher success rate means that the orebodies can be targeted with smaller prospective areas. Figure 13(b) depicts the success-rate curves of the five prospectivity models. It should be noted that the CNN model uses the least prospective areas to target a similar number of orebodies. This indicates that the CNN model, in comparison to the models using hand-crafted variables, has higher targeting efficiency for identifying prospective areas.

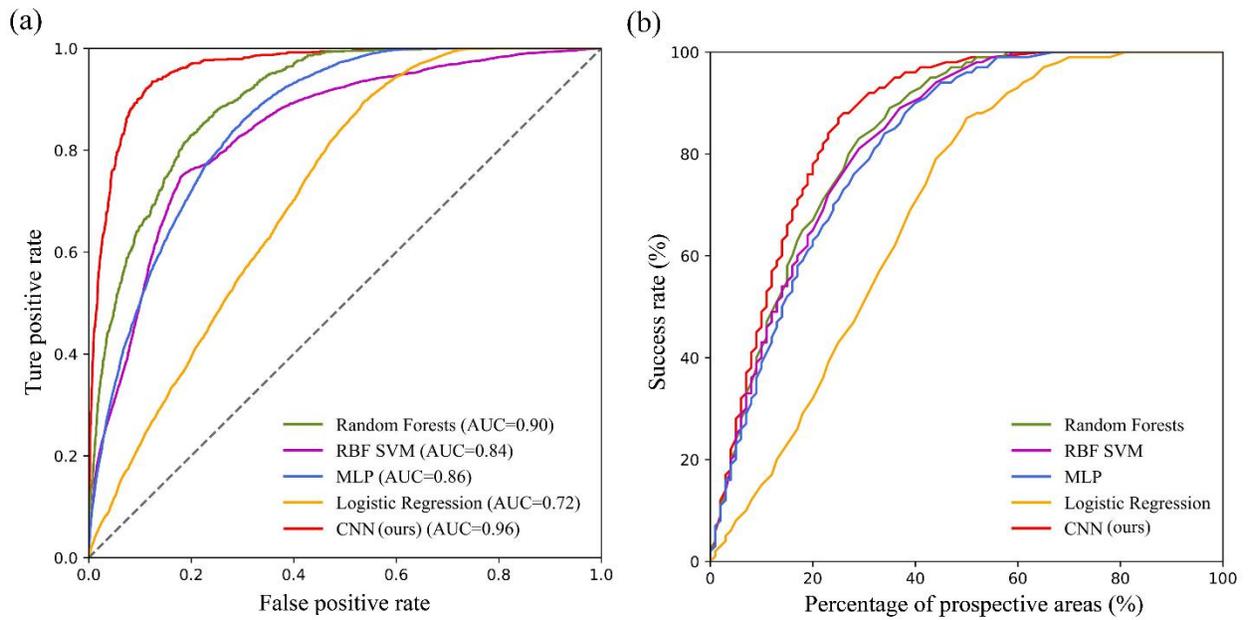

Figure 13: ROC curves (a) and success-rate curves (b) for the five evaluated prospectivity models. The CNN model exhibits the best performance among the five tested methods.

The performance of the CNN model could be impacted by class imbalances between ore-bearing and non-ore-bearing samples. The focal loss in Equation (9) can mitigate this issue. Figure 14 compares the performance of CNN models trained with focal loss and cross entropy. Despite the fact that the class-imbalance problem in our scenario is not severe (the ratio of ore-bearing to non-ore-bearing samples is roughly 1:4), both the AUC and the success rates of our CNN model can be improved by using the focal loss. In the focal loss of the final predictive results, the averaged modulating factors between ore-bearing and non-ore-bearing samples are 0.100 and 0.019, respectively, which demonstrates that focal loss can effectively enhance the focus on the ore-bearing samples in CNN training.

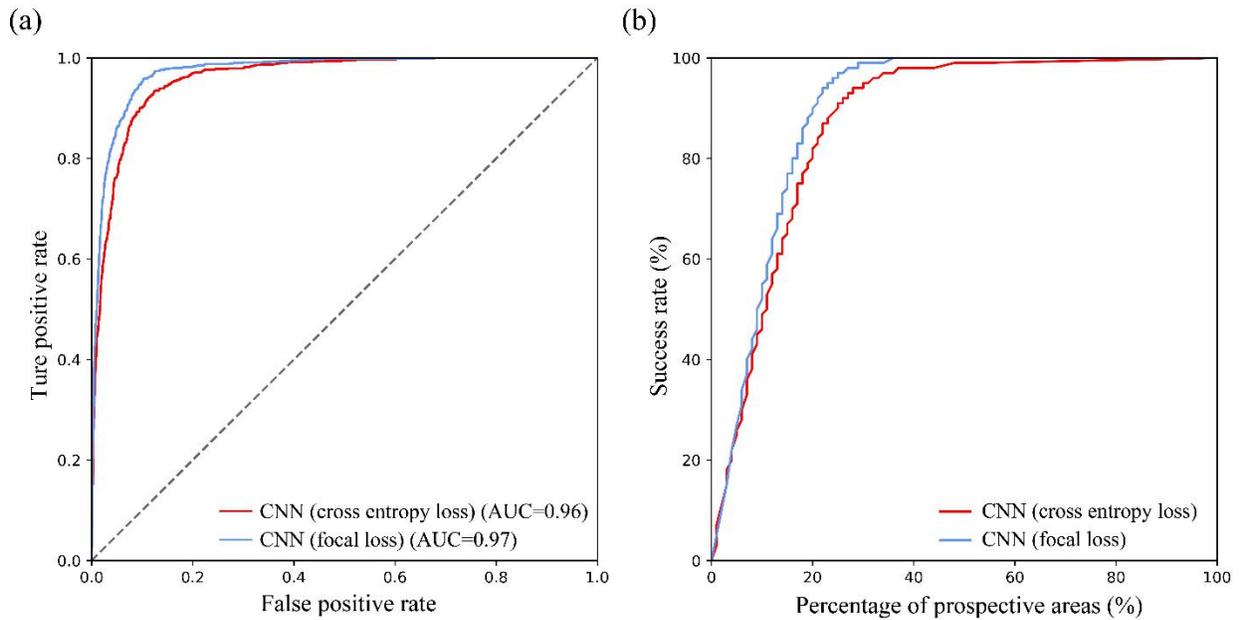

Figure 14: ROC curves (a) and success-rate curves (b) for the CNN models trained by using the cross entropy loss and the focal loss.

4.4. 3D predictive mapping and target appraisal

The high-prospective areas were identified as the potential mineral areas for future exploration. The maximum Youden index (MYI) (Ruopp et al., 2008) was adopted to select an optimum threshold for separating the high-prospective areas from the low-prospective background. We compared the high-prospective areas resulting from the CNN with those from the RF model, which has the best AUC and the success-rate curve among the four prospectivity models that use hand-crafted variables. Table 2 lists the MYIs and the corresponding thresholds for the CNN and the RF model, both of whose thresholds are larger than the prior probability of ore-bearing voxels. Figures 15 shows the high-prospective areas resulting from the CNN and the RF model. The high-prospective areas resulting from the CNN focus on several potential exploration targets. Most of these areas overlap with the high-prospective areas from the RF model, which reflects the fact that the CNN-based model can identify high-prospective areas without being given predictor variables. In comparison, the high-prospective areas identified by the RF model cover more areas and are relatively less discriminative when identifying exploration targets.

Table 2: MYIs and the corresponding thresholds for the CNN and the RF model.

| Prospectivity model | MYI | Probability |
| --- | --- | --- |
| CNN | 0.801 | 0.253 |
| Random forest | 0.569 | 0.394 |

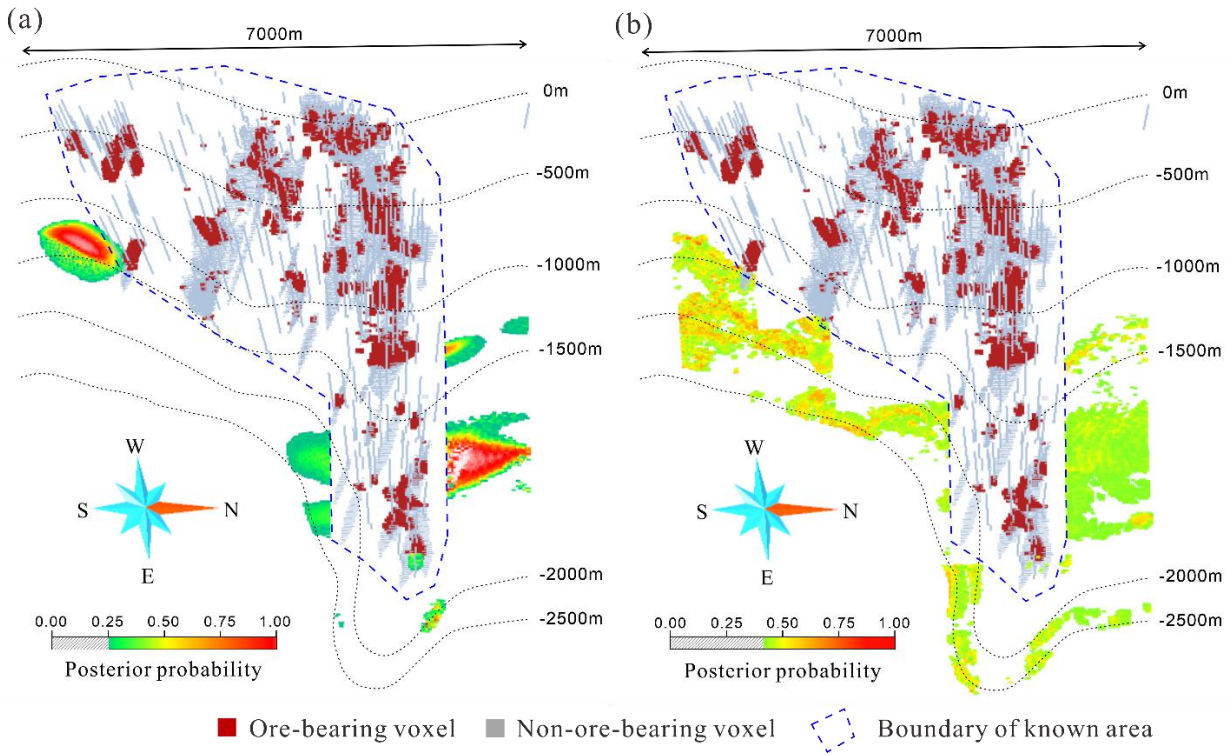

Figure 15: The high-prospective areas resulting from the CNN (a) and the RF (b) model.

According to the predictive results of the CNN model, five mineral exploration targets were identified in the deep-seated areas of Dayingezhuang (Figure 16). The targets I and III extend from the middle of the No. 2 orebodies, at average elevations of –1400 m and –1700 m, respectively, which are considered to be the two horizontal branches of the No. 2 orebodies. Target II is located at 1300 m to 1600 m beneath the Caojiawa orebody, which is considered to be the NE branch of the orebody. Target IV is sited on the NE trending direction of No. 2 orebodies, at depths of 1800 m to 2000 m, which is possibly a deep extension of the No. 2 orebodies. Target V is located SSW of the No. 2 orebodies, at depths of 1900 m to 2300 m, which is the lowest prospective value among the five targets.

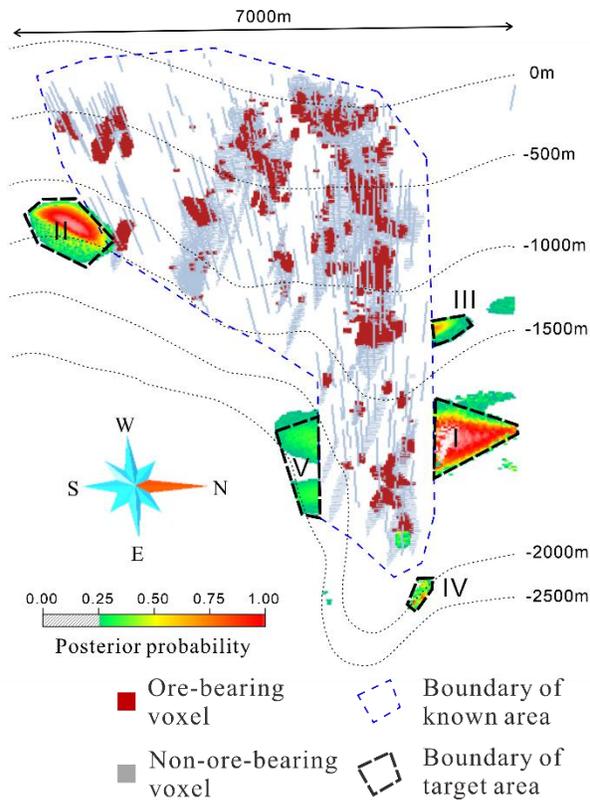

Figure 16: Five mineral exploration targets identified by the CNN model.

## 5. Conclusions

This paper proposes a 3D MPM method that uses CNNs, wherein the 3D mineral prospectivity is directly learned from 3D models. As demonstrated in an application to the Dayingezhuang deposit, a structure-controlled hydrothermal gold deposit, the proposed method has several advantages in comparison to the existing methods for 3D MPM. Firstly, owing to the representation learning ability of CNNs, our method can automatically assess underlying factors affecting mineralization and learn the prospectivity-informative representation for 3D MPM. This allows us to circumvent the need for designing predictor variables in the conventional MPM framework, providing an efficient method to achieve 3D MPM. Secondly, by reorganizing the geometry of the unstructured 3D models into the CNN, the proposed method is capable of exploiting the deep network architecture to learn complex correlations between the geometry of the geological boundary and the localization of mineralization. This leads to high performance in terms of both prediction accuracy and targeting efficiency of 3D MPM. Finally, by learning mineral prospectivity from 3D models, the prospectivity model can be more objective to represent the mineralization control from the geological boundaries, which permits us to prevent the impacts of subjective geological analyses and flawed conceptual models in 3D MPM. These advantages, therefore, make the proposed method a novel but powerful tool to reduce the workload and prospecting risk when targeting deep-seated orebodies.

However, the proposed method has certain limitations. Despite the fact that the CNN model exhibits promising performance, our method only learns mineral prospectivity from a single 3D model owing to the special geological settings in the study area. The proposed network architecture should extend to a more general scenario of multiple ore-controlling geological boundaries. The proposed network architecture can be expanded by reusing it to learn high-level representations from every single 3D model and aggregating the high-level representations for the final MPM. In addition, it is worth noting that, despite the proposed method exhibiting a superior learning capability, its prediction power is dependent on the accuracy of the 3D geological models from which the prospectivity model is learned (see Appendix II in the supplementary material). Analogously to conventional machine learning methods for 3D MPM, adopting 3D models with considerable accuracy and certainty is also essential for identifying reliable exploration targets. In study areas with low data availability, more effort should focus on building multiple plausible 3D models (Lindsay et al., 2013; Wellmann et al., 2018); in this way, learning multiple prospectivity models can reduce the uncertainty of the exploration targets.

## Acknowledgments

This study is partially supported by National Natural Science Foundation of China (Nos. 41972309, 42030809, and 42072325) and National Key Research and Development Program of China (No. 2017YFC0601503).

## Code availability section

3DMPM

Contact: chengyann1222@gmail.com

Hardware requirements: GeForce RTX 1050Ti or higher.

Program language: MATLAB, C++ and Python.

Software required: MATLAB, Visual Studio, Ubuntu and Anaconda.

Program size: 9.67 MB.

The source codes are available for downloading at the link:

https://github.com/ChengYeung1222/3DMPM

List of Figures

13. Figure 13: ROC curves (a) and success-rate curves (b) for the five evaluated prospectivity models. The CNN model exhibits the best performance among the five tested methods.

14. Figure 14: ROC curves (a) and success-rate curves (b) for the CNN models trained by using the cross entropy loss and the focal loss.

15. Figure 15: The high-prospective areas resulting from the CNN (a) and the RF (b) model.

16. Figure 16: Five mineral exploration targets identified by the CNN model.

List of Tables